\documentclass[twocolumn,tighten]{aastex63}
\usepackage{CJKutf8}
\usepackage{amsmath}
\usepackage{booktabs,pbox}
\usepackage{times}

\graphicspath{{./figures/}{./figures/sightlines/}}

\newcommand{\CO}[2]{\mbox{$\mathrm{CO}\,(#1\text{--}#2)$}}

\newcommand{\sbsc}[1]{_\mathrm{#1}}

\newcommand{\ICO}{I\sbsc{CO}}

\newcommand{\sigCO}{\sigma\sbsc{CO}}

\newcommand{\alphaCO}{\alpha\sbsc{CO}}
\newcommand{\Sigmol}{\Sigma\sbsc{mol}}
\newcommand{\sigmol}{\sigma\sbsc{mol}}

\newcommand{\Pturb}{P\sbsc{turb}}
\newcommand{\alphavir}{\alpha\sbsc{vir}}
\newcommand{\rhomol}{\rho\sbsc{mol}}

\newcommand{\Reff}{R\sbsc{eff}}
\newcommand{\rgal}{r\sbsc{gal}}

\newcommand{\uIco}{\mbox{$\rm K\,km\,s^{-1}$}}
\newcommand{\uV}{\mbox{$\rm km\,s^{-1}$}}
\newcommand{\uM}{\mbox{$\rm M_{\odot}$}}
\newcommand{\uSig}{\mbox{$\rm M_{\odot}\,pc^{-2}$}}

\newcommand{\uP}{\mbox{$k_\mathrm{B}\rm\,K\,cm^{-3}$}}

\shorttitle{Molecular Clouds across the Local Galaxy Population}
\shortauthors{SUN et PHANGS}

\begin{document}


\title{Molecular Gas Properties on Cloud Scales Across the Local Star-forming Galaxy Population}



\newcommand{\OSU}{\affil{Department of Astronomy, The Ohio State University, 140 West 18th Avenue, Columbus, OH 43210, USA}}
\newcommand{\Alberta}{\affil{Department of Physics, University of Alberta, Edmonton, AB T6G 2E1, Canada}}
\newcommand{\ANU}{\affil{Research School of Astronomy and Astrophysics, Australian National University, Canberra, ACT 2611, Australia}}
\newcommand{\Caltech}{\affil{Infrared Processing and Analysis Center (IPAC), California Institute of Technology, Pasadena, CA 91125, USA}}
\newcommand{\Carnegie}{\affil{Observatories of the Carnegie Institution for Science, 813 Santa Barbara Street, Pasadena, CA 91101, USA}}
\newcommand{\CCAPP}{\affil{Center for Cosmology and Astroparticle Physics, The Ohio State University, \\191 West Woodruff Avenue, Columbus, OH 43210, USA}}
\newcommand{\CfA}{\affil{Harvard-Smithsonian Center for Astrophysics, 60 Garden Street, Cambridge, MA 02138, USA}}
\newcommand{\CITEVA}{\affil{Centro de Astronomía (CITEVA), Universidad de Antofagasta, Avenida Angamos 601, Antofagasta, Chile}}
\newcommand{\CNRS}{\affil{CNRS, IRAP, 9 Av. du Colonel Roche, BP 44346, F-31028 Toulouse cedex 4, France}}
\newcommand{\ESO}{\affil{European Southern Observatory, Karl-Schwarzschild Stra{\ss}e 2, D-85748 Garching bei M\:{u}nchen, Germany}}
\newcommand{\Heidelberg}{\affil{Astronomisches Rechen-Institut, Zentrum f\"{u}r Astronomie der Universit\"{a}t Heidelberg, \\M\"{o}nchhofstra\ss e 12-14, D-69120 Heidelberg, Germany}}
\newcommand{\IRAM}{\affil{Institut de Radioastronomie Millim\'{e}trique (IRAM), 300 Rue de la Piscine, F-38406 Saint Martin d'H\`{e}res, France}}
\newcommand{\ITA}{\affil{Universit\"{a}t Heidelberg, Zentrum f\"{u}r Astronomie, Institut f\"{u}r Theoretische Astrophysik, \\Albert-Ueberle-Str 2, D-69120 Heidelberg, Germany}}
\newcommand{\IWR}{\affil{Universit\"{a}t Heidelberg, Interdisziplin\"{a}res Zentrum f\"{u}r Wissenschaftliches Rechnen, \\Im Neuenheimer Feld 205, D-69120 Heidelberg, Germany}}
\newcommand{\Maryland}{\affil{Department of Astronomy, University of Maryland, College Park, MD 20742, USA}}
\newcommand{\MPE}{\affil{Max-Planck-Institut f\"{u}r extraterrestrische Physik, Giessenbachstra{\ss}e 1, D-85748 Garching, Germany}}
\newcommand{\MPIA}{\affil{Max-Planck-Institut f\"{u}r Astronomie, K\"{o}nigstuhl 17, D-69117, Heidelberg, Germany}}
\newcommand{\Nagoya}{\affil{Department of Physics, Nagoya University, Furo-cho, Chikusa-ku, Nagoya, Aichi 464-8602, Japan}}
\newcommand{\OAN}{\affil{Observatorio Astron\'{o}mico Nacional (IGN), C/ Alfonso XII, 3, E-28014 Madrid, Spain}}
\newcommand{\ObsParis}{\affil{Sorbonne Universit\'{e}, Observatoire de Paris, Universit\'{e} PSL, CNRS, LERMA, F-75014, Paris, France}}
\newcommand{\Princeton}{\affil{Department of Astrophysical Sciences, Princeton University, Princeton, NJ 08544 USA}}
\newcommand{\Toulouse}{\affil{Universit\'{e} de Toulouse, UPS-OMP, IRAP, F-31028 Toulouse cedex 4, France}}
\newcommand{\UBonn}{\affil{Argelander-Institut f\"ur Astronomie, Universit\"at Bonn, Auf dem H\"ugel 71, 53121 Bonn, Germany}}
\newcommand{\UChile}{\affil{Departamento de Astronom\'{i}a, Universidad de Chile, Camino del Observatorio 1515, Las Condes, Santiago, Chile}}
\newcommand{\UCSD}{\affil{Center for Astrophysics and Space Sciences, Department of Physics, University of California, San Diego, \\9500 Gilman Drive, La Jolla, CA 92093, USA}}
\newcommand{\UGent}{\affil{Sterrenkundig Observatorium, Universiteit Gent, Krijgslaan 281 S9, B-9000 Gent, Belgium}}
\newcommand{\ULyon}{\affil{Univ Lyon, Univ Lyon 1, ENS de Lyon, CNRS, Centre de Recherche Astrophysique de Lyon UMR5574, \\F-69230 Saint-Genis-Laval, France}}
\newcommand{\UMass}{\affil{Department of Astronomy, University of Massachusetts Amherst, 710 North Pleasant Street, Amherst, MA 01003, USA}}
\newcommand{\UWyoming}{\affil{Department of Physics and Astronomy, University of Wyoming, Laramie, WY 82071, USA}}

\author[0000-0003-0378-4667]{\begin{CJK*}{UTF8}{} Jiayi~Sun ({\CJKfamily{gbsn}孙嘉懿}) \end{CJK*}}
\OSU

\author[0000-0002-2545-1700]{Adam~K.~Leroy}
\OSU

\author[0000-0002-3933-7677]{Eva~Schinnerer}
\MPIA

\author[0000-0002-9181-1161]{Annie~Hughes}
\CNRS
\Toulouse

\author[0000-0002-5204-2259]{Erik~Rosolowsky}
\Alberta

\author[0000-0002-0472-1011]{Miguel~Querejeta}
\ESO
\OAN

\author{Andreas~Schruba}
\MPE

\author[0000-0001-9773-7479]{Daizhong~Liu}
\MPIA

\author[0000-0002-2501-9328]{Toshiki~Saito}
\MPIA

\author[0000-0001-6405-0785]{Cinthya~N.~Herrera}
\IRAM

\author[0000-0001-5310-467X]{Christopher~Faesi}
\MPIA
\UMass

\author[0000-0003-1242-505X]{Antonio~Usero}
\OAN

\author[0000-0003-3061-6546]{J\'er\^ome~Pety}
\IRAM
\ObsParis

\author[0000-0002-8804-0212]{J.~M.~Diederik~Kruijssen}
\Heidelberg

\author[0000-0002-0509-9113]{Eve~C.~Ostriker}
\Princeton

\author[0000-0003-0166-9745]{Frank~Bigiel}
\UBonn

\author[0000-0003-4218-3944]{Guillermo~A.~Blanc}
\Carnegie
\UChile

\author[0000-0002-5480-5686]{Alberto~D.~Bolatto}
\Maryland

\author[0000-0003-0946-6176]{M\'{e}d\'{e}ric~Boquien}
\CITEVA

\author[0000-0002-5635-5180]{M\'{e}lanie~Chevance}
\Heidelberg

\author[0000-0002-5782-9093]{Daniel~A.~Dale}
\UWyoming

\author[0000-0003-1943-723X]{Sinan~Deger}
\Caltech

\author[0000-0002-6155-7166]{Eric~Emsellem}
\ESO
\ULyon

\author[0000-0001-6708-1317]{Simon~C.~O.~Glover}
\ITA

\author[0000-0002-3247-5321]{Kathryn~Grasha}
\ANU

\author[0000-0002-9768-0246]{Brent~Groves}
\ANU

\author[0000-0001-9656-7682]{Jonathan~Henshaw}
\MPIA

\author[0000-0002-9165-8080]{Maria~J.~Jimenez-Donaire}
\OAN
\CfA

\author[0000-0002-0432-6847]{Jenny~J.~Kim}
\Heidelberg

\author[0000-0002-0560-3172]{Ralf~S.~Klessen}
\ITA
\IWR

\author[0000-0001-6551-3091]{Kathryn~Kreckel}
\Heidelberg

\author[0000-0002-2278-9407]{Janice~C.~Lee}
\Caltech

\author[0000-0002-6118-4048]{Sharon~Meidt}
\UGent

\author[0000-0002-4378-8534]{Karin~Sandstrom}
\UCSD

\author[0000-0002-5783-145X]{Amy~E.~Sardone}
\OSU
\CCAPP

\author[0000-0003-4161-2639]{Dyas~Utomo}
\OSU

\author[0000-0002-0012-2142]{Thomas~G.~Williams}
\MPIA


\newcommand{\ngalall}{70}
\newcommand{\nlosall}{102\,778}
\newcommand{\ngal}{66}
\newcommand{\nlos}{102\,295}
\newcommand{\ngalhresall}{35}
\newcommand{\nloshresall}{79\,840}
\newcommand{\ngalhres}{32}
\newcommand{\nloshres}{79\,156}
\newcommand{\nlossub}{40\,641}
\newcommand{\nlosdisk}{99\,765}
\newcommand{\nlosdiskhres}{76\,500}
\newcommand{\ngalbar}{43}
\newcommand{\nlosbar}{1\,715}
\newcommand{\ngalnobar}{13}
\newcommand{\ngalucbar}{10}
\newcommand{\ngalarm}{28}

\begin{abstract}
Using the PHANGS-ALMA CO$\,(2\text{--}1)$ survey, we characterize molecular gas properties on ${\sim}100$~pc scales across $102\,778$ independent sightlines in 70 nearby galaxies.
This yields the best synthetic view of molecular gas properties on cloud scales across the local star-forming galaxy population obtained to date. 
Consistent with previous studies, we observe a wide range of molecular gas surface densities (3.4~dex), velocity dispersions (1.7~dex), and turbulent pressures (6.5~dex) across the galaxies in our sample.
Under simplifying assumptions about sub-resolution gas structure, the inferred virial parameters suggest that the kinetic energy of the molecular gas typically exceeds its self-gravitational binding energy at ${\sim}100$~pc scales by a modest factor (1.3 on average).
We find that the cloud-scale surface density, velocity dispersion, and turbulent pressure 
(1) increase towards the inner parts of galaxies,
(2) are exceptionally high in the centers of barred galaxies (where the gas also appears less gravitationally bound), and
(3) are moderately higher in spiral arms than in inter-arm regions.
The galaxy-wide averages of these gas properties also correlate with the integrated stellar mass, star formation rate, and offset from the star-forming main sequence of the host galaxies. These correlations persist even when we exclude regions with extraordinary gas properties in galaxy centers, which contribute significantly to the inter-galaxy variations.
Our results provide key empirical constraints on the physical link between molecular cloud populations and their galactic environment.\\~\\
\end{abstract}


\section{Introduction} \label{sec:intro}

\defcitealias{Sun_etal_2018}{S18}

Observations indicate that the physical properties of giant molecular clouds (GMCs) vary systematically with their location in a galaxy. This result is obtained in the Milky Way \citep[e.g.,][but see \citealt{Lada_Dame_2020}]{Rice_etal_2016,Roman-Duval_etal_2016,Miville-Deschenes_etal_2017,Colombo_etal_2019} and in other galaxies \citep[e.g.,][]{DonovanMeyer_etal_2013,Hughes_etal_2013a,Colombo_etal_2014a,Leroy_etal_2016,Schruba_etal_2019}.
This suggests that GMCs are connected to their galactic context, which affects their formation, structure, or evolution \citep[see, e.g.,][]{Field_etal_2011,Hughes_etal_2013a,Jeffreson_Kruijssen_2018,Meidt_etal_2018,Meidt_etal_2020,Schruba_etal_2019,Sun_etal_2020}.

Understanding this cloud--environment connection has been a challenge because it requires comprehensive, observationally expensive mapping of GMC demographics across the local galaxy population.
This challenge is being addressed by PHANGS-ALMA\footnote{``\underline{P}hysics at \underline{H}igh \underline{A}ngular resolution in \underline{N}earby \underline{G}alaxie\underline{S} with the \underline{A}tacama \underline{L}arge \underline{M}illimeter \underline{A}rray.'' For more information, see \url{www.phangs.org}.}, a large \CO21 line survey covering essentially all ALMA-visible, nearby, massive, star-forming galaxies (A.~K.\ Leroy et al. 2020a, in preparation).
PHANGS-ALMA well samples the local star-forming main sequence across two decades in stellar mass ($10^9{-}10^{11}\,\uM$).
The high resolution and sensitivity of the PHANGS-ALMA data offer an unprecedented opportunity to characterize molecular gas properties on $50{-}150$~pc ``cloud scales'', and the cloud--environment connection across typical star-forming environments in the local universe.

In this Letter, we report measurements of the cloud-scale molecular gas surface density and velocity dispersion, as well as estimates of the turbulent pressure and the virial parameter.
Following our analysis of the PHANGS-ALMA pilot sample of $11$ galaxies \citep[hereafter \citetalias{Sun_etal_2018}]{Sun_etal_2018}, we derive these measurements on fixed $90$~pc and $150$~pc scales using the full PHANGS-ALMA survey, which increases our sample size to \ngalall\ galaxies.
The derived measurements constitute a benchmark data set that can be readily compared with observations of other types of galaxies or numerical simulations reaching similar spatial resolutions \citep[e.g.,][]{Semenov_etal_2018,Dobbs_etal_2019,Fujimoto_etal_2019,Jeffreson_etal_2020}.


\section{Data and Measurements} \label{sec:method}

\textit{Overview:} We carry out a pixel-by-pixel analysis of molecular gas properties at fixed $90$~pc and $150$~pc scales. 
This method provides a simple, reproducible characterization of all detected emission \citep[e.g.,][]{Sawada_etal_2012,Hughes_etal_2013b,Leroy_etal_2016}.
Complementary analyses decomposing the same CO data into individual objects (E.~Rosolowsky et al. 2020, in preparation; A.~Hughes et al., in preparation) yield qualitatively similar conclusions.

\textit{Galaxy sample:}
We include the \ngalall\ PHANGS-ALMA galaxies that had fully-processed ALMA data
by December 2019\footnote{Internal data release \mbox{v3.4}.}.
They consist of $67$ out of the $74$ galaxies in the ALMA Large Program and pilot samples and three nearby galaxies from the extended PHANGS-ALMA sample.
Table~\ref{tab:sample} lists the galaxy sample together with their global properties (columns 1--9).

\textit{Data characteristics:}
The PHANGS-ALMA \CO21 data have native spatial resolutions of $50{-}150$~pc at the distances of the target galaxies, and $1\sigma$ noise levels of $0.2{-}0.3$~K per $2.5$~\uV\ channel.
They combine ALMA interferometric array and single dish observations to recover emission across the full range of spatial scales (see A.~K.\ Leroy et al. 2020b, in preparation).

\textit{Data homogenization:} We convolve the data cubes to a common $150$~pc spatial resolution to allow direct comparison between all \ngalall\ galaxies. 
For a subset of \ngalhresall\ galaxies, we are also able to convolve the cubes to $90$~pc resolution to investigate trends with spatial resolution \citepalias[see also][]{Sun_etal_2018}.

\textit{Product creation:}
We mask the data cubes to only include voxels that contain emission detected with high confidence\footnote{For the release that we use, the masks begin with all regions with $\text{S/N}>3.5$ in three consecutive channels. These masks are then expanded to include all adjacent regions with $\text{S/N}>2$ in two successive channels. The \texttt{Python} realization of this signal identification scheme is available at \url{https://github.com/astrojysun/SpectralCubeTools}.}. We integrate the masked cubes along the spectral axis to produce the integrated intensity ($\ICO$) and effective line width ($\sigCO$) maps.
The latter quantity is derived as $\sigCO = \ICO / (\sqrt{2\pi}\, T_\mathrm{peak})$ following \citet{Heyer_etal_2001}\footnote{Note that for a Gaussian line profile, $\sigCO$ equals its dispersion.}, where $T_\mathrm{peak}$ is the brightness temperature at the line peak, and is subsequently corrected for the instrumental line broadening following \citetalias{Sun_etal_2018} (see equation~5 therein).
We produce associated uncertainty maps via error propagation from the estimated noise in the cube.
This product creation scheme closely follows \citetalias{Sun_etal_2018} and is detailed in A.~K.~Leroy et al.\ (2020b, in preparation).

Our masking scheme guarantees high S/N ratio CO line measurements at the expense of excluding faint CO emission, especially from sightlines with low $\ICO$ and high $\sigCO$.
The resultant data censoring function is shown in Figure~\ref{fig:overview} (see formula in Appendix~\ref{apdx:censoring}).
We report in Table~\ref{tab:sample} (column~12) the CO flux completeness for each galaxy (the flux within the mask divided by the total flux in the data cube, or $f\sbsc{CO}$).

\textit{Resampling:} We resample the two-dimensional maps of $\ICO$, $\sigCO$, and their uncertainties with hexagonal pixels matching the beam size.
This ensures that the resampled measurements are nearly mutually independent.
We list the number of independent measurements (sightlines) in each galaxy in Table~\ref{tab:sample} (column~13).

\textit{Conversion to physical quantities:}
We use $\sigCO$ as a tracer of the molecular gas velocity dispersion, $\sigmol$.
We derive molecular gas surface density, $\Sigmol$, via
\begin{equation}
    \Sigmol = \alphaCO\;R_{21}^{-1}\;\ICO~. \label{eq:Sigmol}
\end{equation}
\noindent Here $R_{21}=0.65$ is the adopted \CO21-to-\CO10 line ratio \citep[]{Leroy_etal_2013b,denBrok_etal_2020b} and $\alphaCO$ is the CO-to-H$_2$ conversion factor. We adopt a metallicity-dependent $\alphaCO$
\citep[similar to the metallicity-dependent part of the xCOLDGASS prescription;][]{Accurso_etal_2017}:
\begin{equation}
    \alphaCO = 4.35\,Z'^{-1.6}\;\uSig\,(\uIco)^{-1}~,
    \label{eq:alphaCO}
\end{equation}
\noindent where $Z'$ refers to the local metallicity in units of the solar value.
Following \citet{Sun_etal_2020}, we estimate $Z'$ based on galaxy global stellar mass and effective radius (see Table~\ref{tab:sample} for values and data sources), assuming a galaxy global mass--metallicity relation \citep{Sanchez_etal_2019} and a fixed radial metallicity gradient within a galaxy \citep{Sanchez_etal_2014}.

We use $\Sigmol$ and $\sigmol$ to estimate the mean turbulent pressure in the molecular gas, $\Pturb\approx\rho\sbsc{mol}\,\sigma\sbsc{mol}^2$, and the virial parameter, $\alphavir$, via
\begin{align}
    \Pturb =\;& 3.3\times10^4\,\uP\;\times \nonumber\\
    &\left(\frac{\Sigma\sbsc{mol}}{10^2\;\uSig}\right) \left(\frac{\sigma\sbsc{mol}}{10\;\uV}\right)^2 \left(\frac{D\sbsc{beam}}{150\;\mathrm{pc}}\right)^{-1}~, \label{eq:P_turb}\\
    \alphavir =\;& 3.1\;\times \nonumber\\
    &\left(\frac{\Sigma\sbsc{mol}}{10^2\;\uSig}\right)^{-1} \left(\frac{\sigma\sbsc{mol}}{10\;\uV}\right)^2 \left(\frac{D\sbsc{beam}}{150\;\mathrm{pc}}\right)^{-1}~. \label{eq:alpha_vir}
\end{align}
\noindent Here $D\sbsc{beam}$ denotes the beam full-width-half-maximum and is assumed to be the depth along the line of sight\footnote{This differs from the assumptions adopted in the complementary cloud identification analysis (E.~Rosolowsky et al. 2020, in preparation): in that paper, a gas cloud's extent along the line of sight is limited to $<100$~pc.}.
Both equations assume a single, spherical gas structure filling each beam \citepalias{Sun_etal_2018}. The second equation assumes in addition that the gas structures have a density profile of $\rho(r) \propto r^{-1}$ \citep[e.g., following][]{Rosolowsky_Leroy_2006}.


\section{Results} \label{sec:result}

We measure $\Sigmol$, $\sigmol$, $\Pturb$, and $\alphavir$ on cloud scales in a homogeneous way across our sample.
This yields \nlosall\ independent measurements at $150$~pc resolution in \ngalall\ galaxies, and \nloshresall\ measurements at $90$~pc resolution in \ngalhresall\ galaxies.
These measurements are published in Table~\ref{tab:product} in a machine-readable form.
We focus on the $150$~pc scale measurements, which are available for all \ngalall\ galaxies, while occasionally referencing to the $90$~pc scale measurements to illustrate resolution dependencies.

In the following data analysis and presentation, we omit measurements from four galaxies (NGC~4207, NGC~4424, NGC~4694, and NGC~4826).
These galaxies have edge-on orientation (NGC~4207) and/or peculiar gas kinematics due to ram-pressure stripping (NGC~4424), a strong nuclear outflow (NGC~4694), or represent a recent merger event (NGC~4826).
We still report results for these galaxies in the tables, but below we focus on the remaining \nlos\ sightlines in \ngal\ galaxies for data presentation.

\subsection{Statistics of Cloud-scale Molecular Gas Properties}\label{sec:result:stats}

\begin{figure*}[htp]
\fig{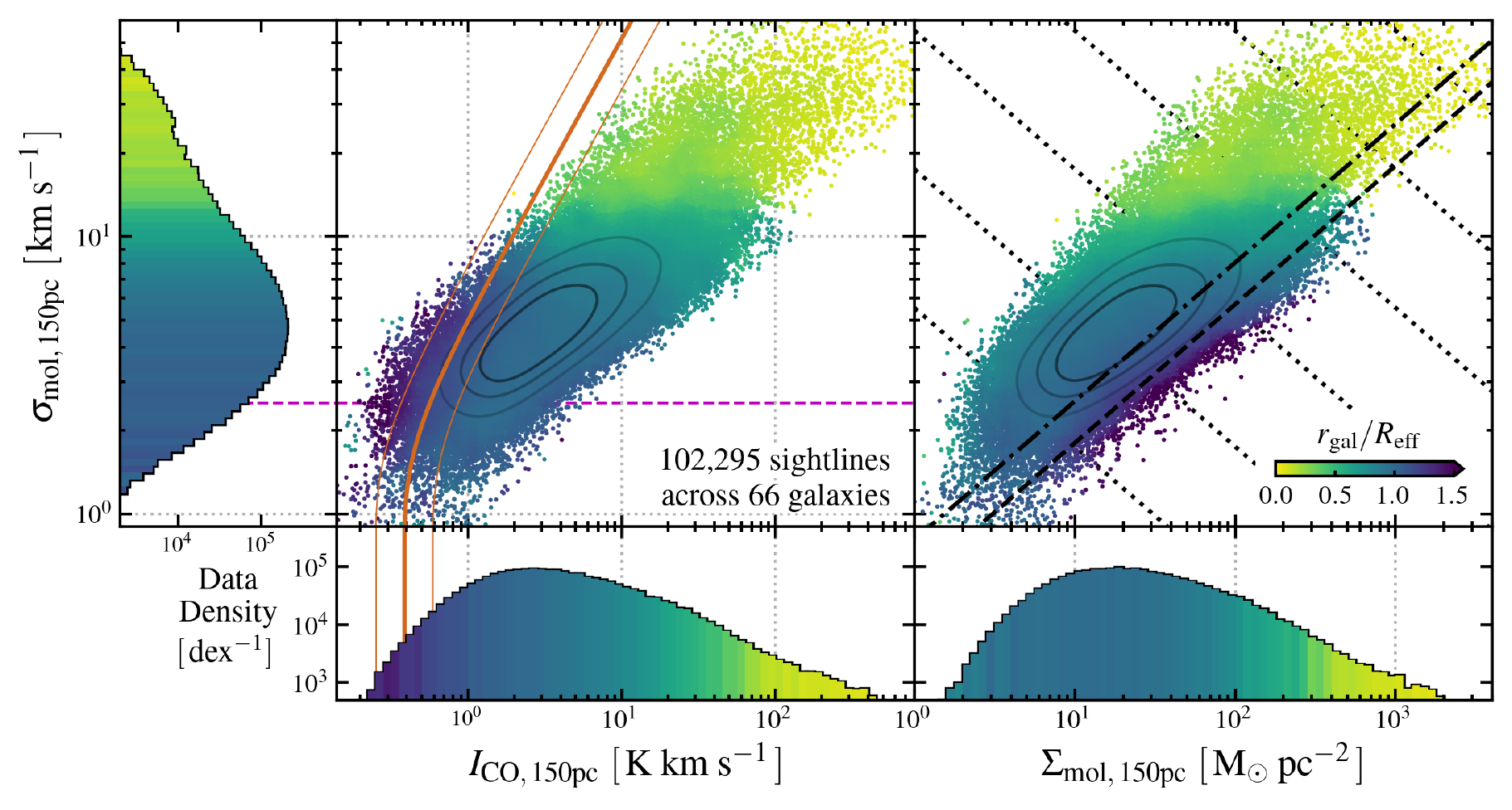}{0.98\textwidth}{}
\vspace{-2.0em}
\caption{
\textbf{Strong and location-dependent variations in CO line intensity ($\ICO$), molecular gas surface density ($\Sigmol$), and velocity dispersion ($\sigmol$) on cloud scales in nearby star-forming galaxies.}
Each of the \nlos\ data points represents an independent measurement on $150$~pc scale.
Data density contours enclose 30, 50, and 70\% of data points.
Color indicates the galactocentric radius $\rgal$ of the measurement normalized by the host galaxy's effective radius $\Reff$. 
Color in the scatter plots represents the median $\rgal/\Reff$ for data points with similar $\ICO$, $\Sigmol$, and $\sigmol$.
Color in the histograms indicates the median $\rgal/\Reff$ in each bin.
The brown lines show the censoring function, to the left of which little CO emission can be detected. As this function varies from galaxy to galaxy, we use the thick line to show the median and the thin lines to show the $1\sigma$ range across all targets.
The magenta dashed line indicates the channel width of the CO observations, above which the $\sigmol$ measurements are most reliable.
The black dashed and dashed-dotted lines in the top-right panel show the expected relation for beam-filling, spherical GMCs with virial parameter $\alphavir=1$ and $2$, respectively. The series of dotted lines in the background show the loci of constant $\Pturb$ (from the lower left to the upper right: $\Pturb=10^3,\,10^4,\,\ldots,\,10^8\,\uP$).
\label{fig:overview}}
\end{figure*}

Figure~\ref{fig:overview} shows the distributions of $\ICO$, $\Sigmol$, and $\sigmol$, measured at $150$~pc resolution. 
This figure encapsulates molecular gas properties on cloud scales across a wide range of star-forming environments at $z \approx 0$.
Table~\ref{tab:stats} provides area-weighted statistics treating each sightline equally, and $\Sigmol$-weighted statistics treating each quantum of molecular gas mass equally.

We observe median $\ICO$, $\Sigmol$, and $\sigmol$ values similar to the results of previous studies \citep[e.g.,][]{Bolatto_etal_2008,Colombo_etal_2019}, but with a large spread.
Across the full sample at $150$~pc resolution, we find mass-weighted median $\Sigmol=110\;\uSig$ and mass-weighted median $\sigmol=9.1\;\uV$ (see Table~\ref{tab:stats}). Given the broad sample selection and coverage, these can be taken as typical values across the local star-forming galaxy population.
We also see that the $\pm3\sigma$ (i.e., $99.7$\%) range of the mass-weighted $\Sigmol$ and $\sigmol$ distributions is large, $3.4$ and $1.7$ dex respectively. Given that data censoring hinders the detection of low $\ICO$ signals (to the left of the brown curves in Figure~\ref{fig:overview}), the true ranges of $\ICO$ and $\Sigmol$ are likely even wider.

We find a strong and statistically significant\footnote {Here and in subsequent sections ``statistically significant'' means $p\text{-value} \ll 0.001$.} correlation between $\Sigmol$ and $\sigmol$ (Spearman's rank correlation coefficient $\rho=0.77$).
This correlation results in even stronger variations in $\Pturb\approx\rhomol\,\sigmol^2$ than those in $\Sigmol$ or $\sigmol$ alone.
Indeed, $\Pturb$ varies by $\gtrsim6$~dex at $\pm3\sigma$ across our sample (Figure~\ref{fig:overview} and Table~\ref{tab:stats}).

We further compare the observed $\Sigmol{-}\sigmol$ distribution to the expected relations for beam-filling, spherical clouds with fixed virial parameters $\alphavir$ (black diagonal lines in Figure~\ref{fig:overview}; see Equation~\ref{eq:alpha_vir}).
These relations capture the overall trend in the data, with the $\alphavir=1$ line lying near their lower envelope.
Across our full sample, $\alphavir$ has a mass-weighted median value of $2.7$ and a $1\sigma$ scatter of $0.7$~dex (Table~\ref{tab:stats}).
This means that the kinetic energy in the molecular gas on average slightly exceeds its gravitational binding energy by a factor of $1.3$ on $150$~pc scales.
This is consistent with the conclusion in \citet{Sun_etal_2020} that the observed molecular gas velocity dispersion at ${\sim}100$~pc scales mainly reflects gas motions due to self-gravity and, to a lesser degree, to external gravity and ambient pressure.

The calculation of $\Pturb$ and $\alphavir$ assumes an idealized sub-beam gas distribution (see Section~\ref{sec:method}).
In reality, the molecular gas remains clumpy on $\lesssim100$~pc spatial scales \citep{Leroy_etal_2013a}, and the small-scale density distribution may vary from place to place.
These variations in sub-beam density distribution may introduce systematic uncertainties in our inferred $\Pturb$ and $\alphavir$ values.
Nevertheless, our $\Sigmol$ and $\sigmol$ measurements should still capture the true distribution of molecular gas properties at the fixed $150$~pc spatial scale.

To further illustrate the effect of resolution on our analysis, we compare our measurements at $150$~pc scales with those at $90$~pc scales for the \ngalhres\ galaxies that have data at both resolutions (see Table~\ref{tab:stats} for the statistics at $90$~pc).
This includes \nloshres\ independent sightlines at $90$~pc scales or \nlossub\ sightlines at $150$~pc scales.
We find the mass-weighted medians of $\ICO$, $\Sigmol$, and $\Pturb$ at $90$~pc scales to be moderately higher than the $150$~pc scale values by factors of $1.5$, $1.5$, and $2.0$, respectively.
However, we see little difference in the median values of $\sigmol$ and $\alphavir$ at the two resolutions, and
the observed dynamic ranges of all quantities is essentially the same at both spatial scales.
This suggests that the gas is moderately clumped below our resolution, but that our qualitative conclusions are not sensitive to resolution-related biases and robustly reflect typical molecular gas properties at ${\sim}100$~pc scales.

\subsection{Correlation with Galactocentric Radius}\label{sec:result:rgal}

\begin{figure*}[thbp]
\gridline{
\fig{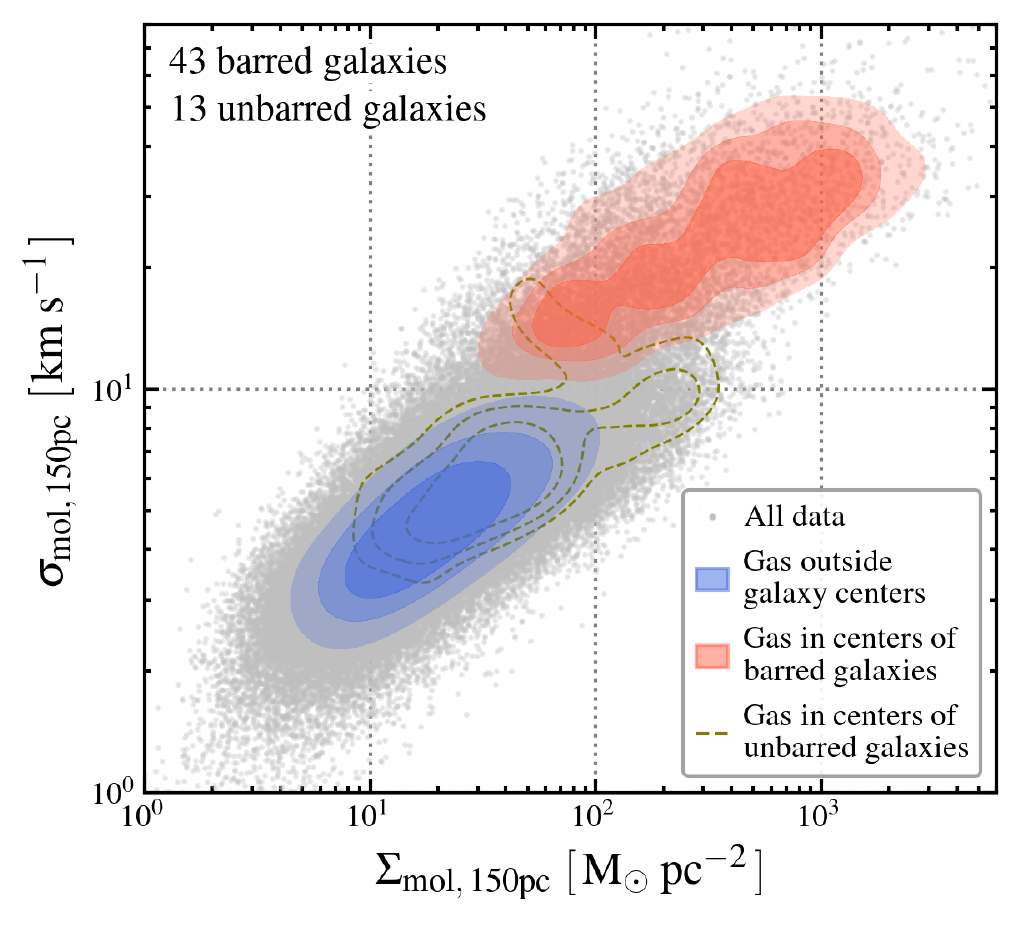}{0.487\textwidth}{}
\fig{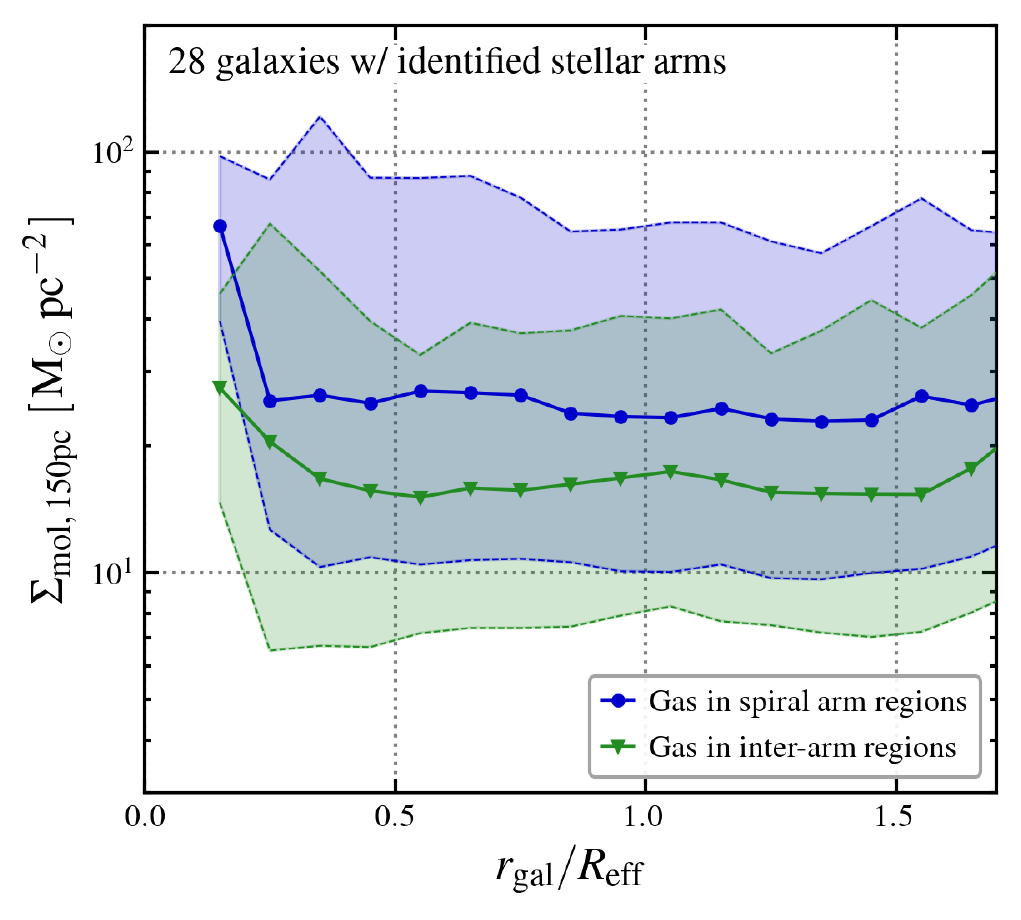}{0.49\textwidth}{}
}
\vspace{-3.0em}
\caption{
\textbf{\textit{Left:} Molecular gas in the centers of barred galaxies shows distinctively high $\Sigmol$ and $\sigmol$.}
Data density contours show the distribution of measurements in galaxy disks (i.e., outside the central regions; blue filled contours), in the centers of \ngalbar\ identified barred galaxies (red filled contours), and in the centers of \ngalnobar\ unbarred galaxies (brown dashed contours).
We see typical gas properties of $\Sigmol\gtrsim100\,\uSig$ and $\sigmol\gtrsim10\,\uV$ in the centers of barred galaxies.
This is in sharp contrast to the gas properties in galaxy disks and in the centers of unbarred galaxies.
\textbf{\textit{Right:} Molecular gas in spiral arms displays higher $\Sigmol$ than the gas in inter-arm regions at given galactocentric radii.}
The blue and green curves show the trends of median $\Sigmol$ in each $\rgal/\Reff$ bin for the gas in spiral arms and inter-arm regions, respectively. Shaded regions denote the $1\sigma$ ($68.3$\%) range of the binned data distribution.
In the \ngalarm\ galaxies exhibiting stellar spiral structures, the gas in the spiral arms typically shows $1.5{-}2$ times higher $\Sigmol$ than the detected gas in the inter-arm regions at the same $\rgal/\Reff$.
\label{fig:regions}}
\end{figure*}

The variations in $\ICO$, $\Sigmol$, and $\sigmol$ correlate with location in the host galaxy.
To illustrate this, we color-code all data points in Figure~\ref{fig:overview} by their galactocentric radii ($\rgal$), normalized to the effective radii of their host galaxies ($\Reff$; see Table~\ref{tab:sample} for values and data sources).
Both $\Sigmol$ and $\sigmol$ tend to increase towards smaller $\rgal/\Reff$.
Additionally, the gas in the inner regions ($\rgal/\Reff<0.5$) frequently shows enhanced $\sigmol$ at a given $\Sigmol$.

These radial trends are partly driven by the structure of galaxy disks.
Most star-forming galaxies show increasing gas and stellar mass surface densities towards their central regions.
This leads to a similar radial trend on the mean pressure in the interstellar medium (ISM) required to keep it in vertical dynamical equilibrium \citep[][]{Elmegreen_1989,Blitz_Rosolowsky_2004,Ostriker_etal_2010}.
We expect the same trend to hold for the turbulent pressure in the molecular gas, $\Pturb$, which correlates with the mean ISM pressure \citep{Sun_etal_2020}.
This expectation matches well with the trend of decreasing $\rgal/\Reff$ with increasing $\Pturb$ values
in Figure~\ref{fig:overview}.

The expectation from ISM dynamical equilibrium does not by itself explain all the trends in Figure~\ref{fig:overview}---for fixed $\Sigmol$, we also find excess $\sigmol$ at smaller $\rgal/\Reff$.
At face value, this suggests that molecular gas in the inner galaxy tends to be more weakly bound (higher $\alphavir$) than the gas in the outer galaxy.
Such a trend is expected from the larger contribution of the external (mostly stellar) potential to the dynamical equilibrium at smaller radii \citep[e.g., \citetalias{Sun_etal_2018};][]{Meidt_etal_2018,Gensior_etal_2020}.
However, the observed trend could instead suggest that the gas is more clumpy in the inner parts of galaxies, or that our prescription over-predicts $\alphaCO$ in the outer disks of galaxies.
\added{If we adopt an alternative prescription with $\alphaCO \propto Z'^{-0.5}$ (as suggested by recent numerical simulations; M.~Gong et al., private communication), the apparent trend of lower $\sigmol$ at fixed $\Sigmol$ towards the outer disks (i.e., $\rgal/\Reff\gtrsim1.5$) would disappear. But the elevated $\sigmol$ at fixed $\Sigmol$ near the galaxy centers (i.e., $\rgal/\Reff\lesssim0.5$) would persist and thus cannot be explained by $\alphaCO$ alone.}

The trend with galactocentric radius at fixed $\Sigmol$ may also reflect biases in the line width measurement. Using the CO rotation curves from \citet{Lang_etal_2020}, we verified that unresolved rotation often represents a minor contribution to our measured line width at $90{-}150$~pc scales in the inner parts of galaxies. However, unresolved non-circular motions may still play an important role \citep[e.g.,][]{Colombo_etal_2014b,Meidt_etal_2018,Meidt_etal_2020,Henshaw_etal_2020}.

\subsection{Correlation with Galaxy Bars and Spiral Arms}\label{sec:result:regions}

We investigate whether galaxy morphological features, i.e., stellar bars and spiral arms, have an impact on the molecular gas properties on cloud scales.
We classify each target galaxy as barred or unbarred (see Table~\ref{tab:sample}), and divide the PHANGS-ALMA CO footprint into a central region and a disk region based on near infrared images.
The central regions often correspond to distinct structures (e.g., nuclear rings) showing extra light at near infrared wavelengths. For galaxies with strong spiral arms, we further identify arm regions and the corresponding inter-arm regions covering the same $\rgal$ range.
The methodology closely follows \citet{Salo_etal_2015} and \citet{Herrera-Endoqui_etal_2015} and is detailed in M.~Querejeta et al.\ (2020, in preparation).

The left panel of Figure~\ref{fig:regions} compares molecular gas properties in the central regions and the disk regions of our galaxies.
Motivated by previous studies \citep[e.g.,][\citetalias{Sun_etal_2018}]{Sakamoto_etal_1999,Jogee_etal_2005}, we indicate the centers of \ngalbar\ galaxies classified as barred and \ngalnobar\ galaxies classified as unbarred separately\footnote{The remaining \ngalucbar\ galaxies have ambiguous classifications (see Table~\ref{tab:sample}). Measurements in their central regions are omitted in Figure~\ref{fig:regions}.}.
The centers of barred galaxies show ${\sim}20$ times higher mass-weighted median $\Sigmol$ and ${\sim}5$ times higher mass-weighted median $\sigmol$ compared to the disk regions (Table~\ref{tab:stats}).
These central regions of barred galaxies mostly host molecular gas with $\Sigmol\gtrsim100\;\uSig$ and $\sigmol\gtrsim10\;\uV$\replaced{.}{and commonly show excess in star formation.}
A small fraction of the gas in the centers of unbarred galaxies also shows high $\Sigmol$ and $\sigmol$, but the majority resembles the gas in the disk regions.
This sharp contrast between barred and unbarred galaxies indicates that the high $\Sigmol$ and $\sigmol$ frequently found in star-forming galaxy centers is linked to the presence of stellar bars.

Our measurements in galaxy centers can be affected by uncertainty related to $\alphaCO$ and $R\sbsc{21}$. \citet{Sandstrom_etal_2013} and \citet{denBrok_etal_2020b} find evidence for low $\alphaCO$ and high $R\sbsc{21}$ in star-forming galaxy centers.
If our prescription also accounted for these effects, the $\Sigmol$ enhancement would be reduced in galaxy centers, but the excess in $\sigmol$ at a given $\Sigmol$ would be even more extreme relative to disks.

The observed extreme gas properties in barred galaxy centers are consistent with existing knowledge about the role of stellar bars in regulating ISM properties.
Stellar bars can drive large-scale gas inflows, boosting the central gas reservoir and leading to high $\Sigmol$ \citep[e.g.,][]{Pfenniger_Norman_1990,Sakamoto_etal_1999,Sheth_etal_2002,Jogee_etal_2005,Tress_etal_2020a}. 
Meanwhile, the released gravitational energy from gas inflow as well as the stronger local stellar and AGN feedback together enhance the local turbulence \citep[e.g.,][]{Kruijssen_etal_2014,Armillotta_etal_2019,Sormani_etal_2019}. Complex gas streaming motions that are unresolved in our data could also bias $\sigmol$ higher than the turbulent velocity dispersion \citep[e.g.,][]{Henshaw_etal_2016}.

The right panel in Figure~\ref{fig:regions} compares the distribution of $\Sigmol$ at fixed $\rgal/\Reff$ in spiral arm regions and inter-arm regions for \ngalarm\ galaxies with identifiable spiral structures in their stellar distribution.
Molecular gas in the arm regions shows typically $1.5{-}2$ times higher $\Sigmol$ relative to the gas in the inter-arm regions at fixed $\rgal/\Reff$.
 We further find (not shown in Figure~\ref{fig:regions}) that the gas in spiral arms shows ${\sim}20\%$ higher $\sigmol$, ${\sim}2{-}3$ times higher $\Pturb$, and ${\sim}15$\% lower $\alphavir$ at fixed $\rgal/\Reff$.
Consistent with previous studies examining individual galaxies \citep[e.g.,][]{Hughes_etal_2013b,Colombo_etal_2014a,Hirota_etal_2018}, these results support the idea that spiral arms harbor more high surface density, turbulent, bound molecular clouds.

Though statistically significant, the measured contrast in $\Sigmol$ between arms and inter-arm gas may seem lower than what one would expect from visual inspection of the PHANGS-ALMA CO maps (e.g., figure~12 in \citetalias{Sun_etal_2018}). There the spiral arms typically appear replete with bright emission, while the inter-arm regions show only sporadic, faint signal with a large portion of the area lacking significant CO detection. We note that our quantitative analysis focuses solely on the gas securely detected in CO without accounting for the area covering fraction of the CO detection.
Had we included map pixels with non- or low-significance detections in our analysis, measurements in the inter-arm regions would be more severely diluted than measurements in the arm regions, and the arm versus inter-arm contrast would be considerably larger than the factor of $1.5{-}2$ measured above (also see M.~Querejeta et al. 2020, in preparation).

To summarize, our measurements based on significant detections of CO emission reveal moderate differences between the molecular gas properties in spiral arm versus inter-arm regions. In addition to this, the spatial density of secure CO detections is much lower in the inter-arm regions than in the spiral arms of galaxies. Together, these two observations suggest that spiral arms not only accumulate molecular gas but also lightly modify the properties of the gas \citep[e.g.,][]{Dobbs_Bonnell_2008,Tress_etal_2020}.

\subsection{Correlation with Integrated Galaxy Properties}\label{sec:result:global}

\begin{figure*}[htbp]
\gridline{
\fig{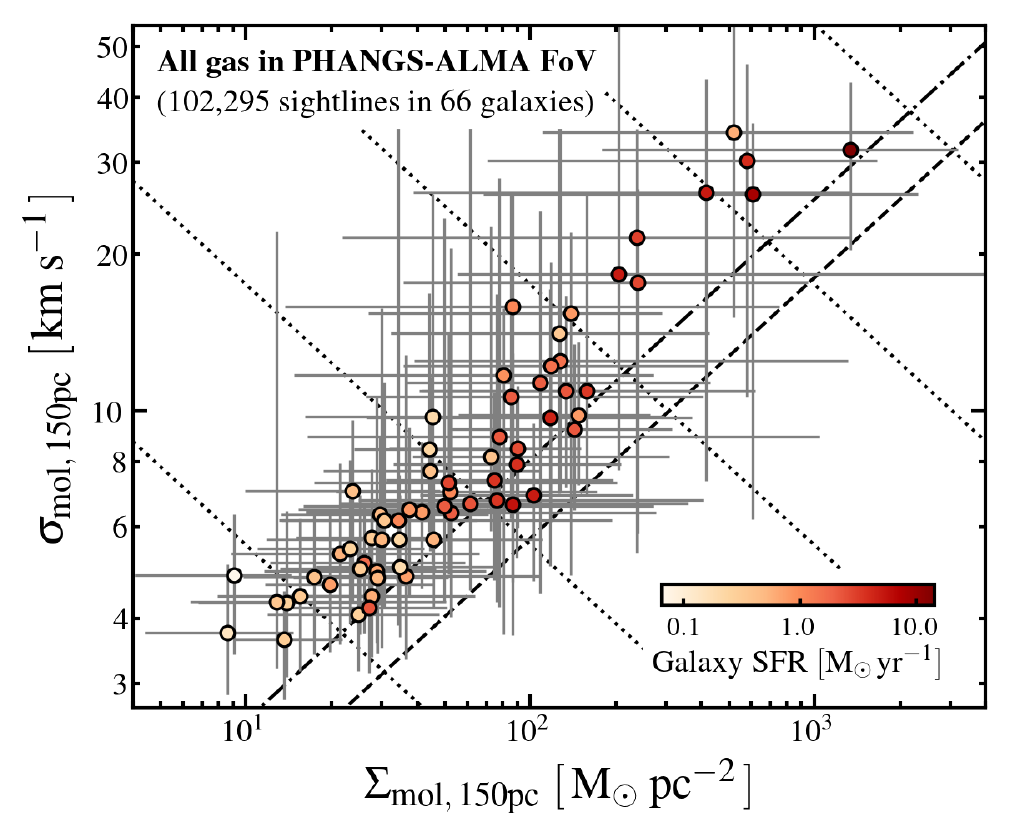}{0.47\textwidth}{}
\fig{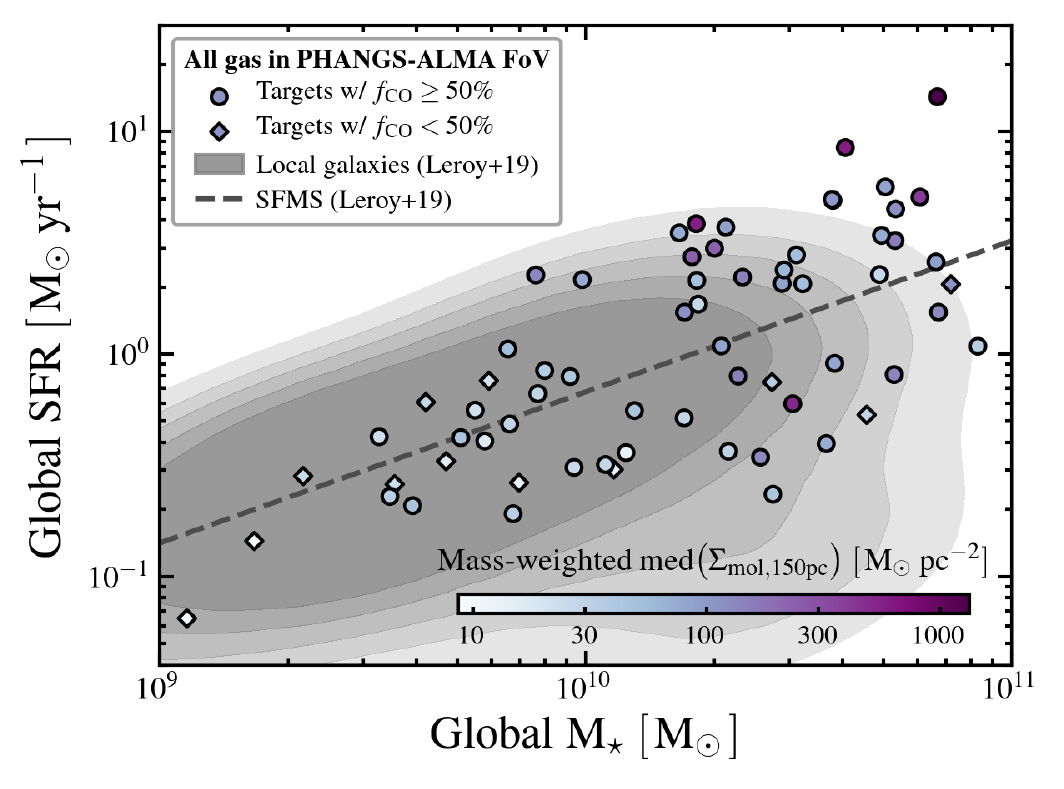}{0.528\textwidth}{}
}
\vspace{-3em}
\gridline{
\fig{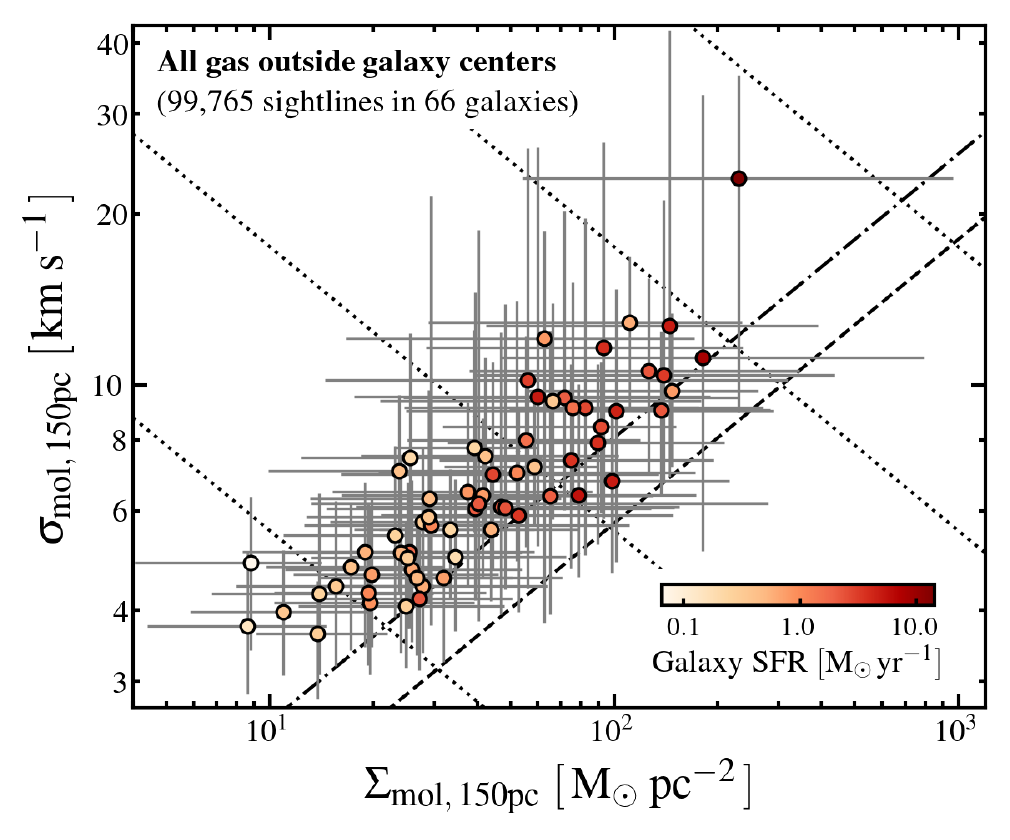}{0.47\textwidth}{}
\fig{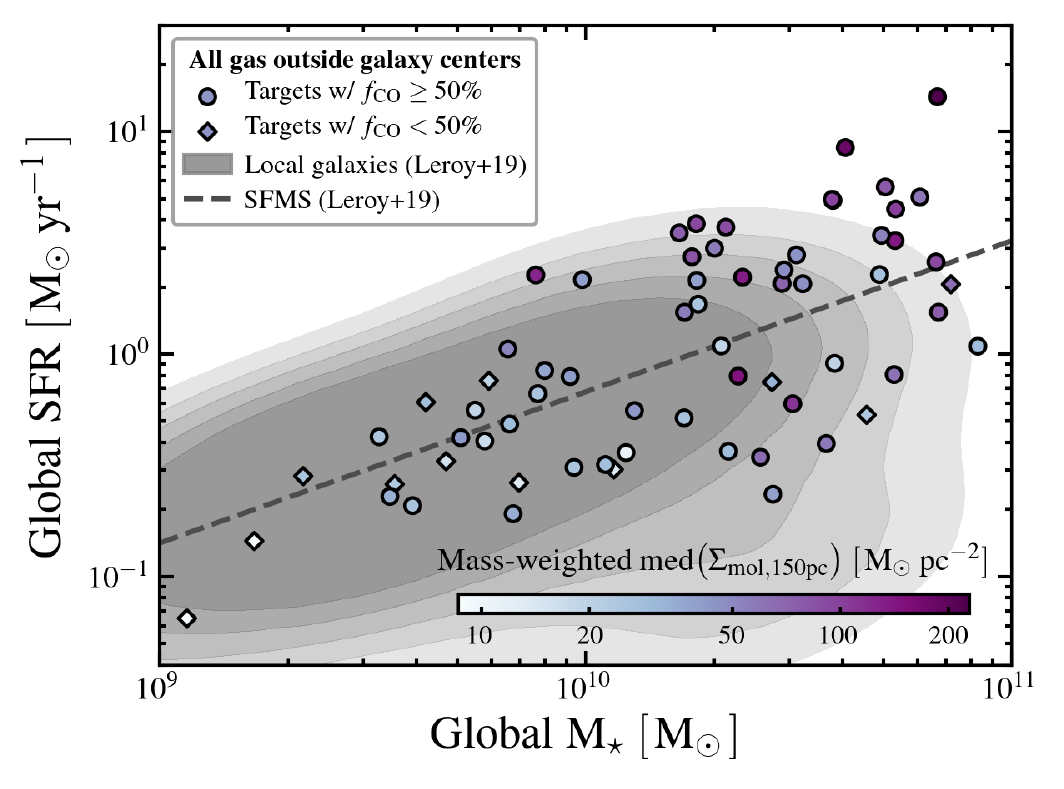}{0.528\textwidth}{}
}
\vspace{-2.5em}
\caption{
\textbf{Molecular gas in more massive and more actively star-forming galaxies shows higher surface densities $\Sigmol$ and velocity dispersions $\sigmol$ on $150$~pc scales.}
\textit{Top left:}
Each point shows the mass-weighted median value of $\Sigmol$ and $\sigmol$ across the PHANGS-ALMA field of view in a galaxy, and the error bars indicate their 1$\sigma$ ($68.3$\%) range.
The diagonal lines represent constant loci of $\alphavir$ and $\Pturb$ as in Figure~\ref{fig:overview}.
Galaxies with a higher global SFR (denoted by darker color) tend to show higher median $\Sigmol$ and $\sigmol$ on cloud scales.
\textit{Top right:}
The \ngal\ galaxies studied here (large symbols) are color-coded by their galaxy-wide, gas mass-weighted median $\Sigmol$ and overlaid on the $M_\star{-}\text{SFR}$ distribution of all local galaxies \citep[gray contours;][]{Leroy_etal_2019}.
Galaxies with low ($<50\%$) CO flux completeness are shown with a different symbol.
The disk-wide median $\Sigmol$ shows significant correlation with galaxy global properties, including stellar mass, SFR, and offset from the star-forming main sequence \citep[gray dashed line;][]{Leroy_etal_2019}.
\textit{Bottom panels}: Similar to the top panels, but with each point showing the statistics for all the gas outside the central regions in each galaxy. The galaxy-to-galaxy variations in gas properties become smaller, but the same general trends with galaxy global properties persist.
The single outlier showing high mass-weighted median $\sigmol$ in the bottom left panel is NGC~1365, for which the PHANGS-ALMA CO map only covers the stellar bar-covered inner part of the galaxy.
\label{fig:MS}}
\end{figure*}

We find that molecular gas properties on cloud scales correlate with integrated properties of the host galaxies.
In Figure~\ref{fig:MS}, the top left panel shows the mass-weighted median $\Sigmol$ and $\sigmol$ values on $150$~pc scale within each galaxy, with each point colored by the galaxy global star formation rate (SFR).
The top right panel shows how the galaxy-wide, mass-weighted median $\Sigmol$ varies among galaxies across the galaxy global $M_\star{-}\text{SFR}$ space.

Across our sample, the mass-weighted median $\Sigmol$ and $\sigmol$ vary by $2$~dex and $1$~dex from galaxy to galaxy, respectively.
These cloud scale gas properties also show statistically significant correlations with host galaxy global $M_\star$ (Spearman's $\rho=0.64$ and $0.53$), global SFR ($\rho=0.72$ and $0.58$), and offset in SFR from the local star-forming main sequence ($\Delta\text{MS}$; $\rho=0.45$ and $0.35$).
We also find positive correlations between the mass-weighted median $\Pturb$ and the same galaxy global properties (not shown in Figure~\ref{fig:MS}).
The mass-weighted median of $\alphavir$, however, shows an anti-correlation with the galaxy's SFR ($\rho=-0.44$)
and $\Delta\text{MS}$ ($\rho=-0.41$). Figure~\ref{fig:MS} only shows the $150$~pc results, but we see similar trends using data at $90$~pc resolution.

The pronounced galaxy-to-galaxy variations in these mass-weighted median quantities is partly explained by galaxies in our sample that host a distinct central concentration of CO-bright molecular gas. This is especially true of barred galaxies, where the central regions host a substantial fraction of the galaxy's molecular gas mass. In these galaxies, the exceptional gas properties in the central region bias the galaxy-wide mass-weighted median measurements toward high $\Sigmol$ and $\sigmol$.
In light of this bias, we also calculate and compare the mass-weighted median properties for all the CO emission \textit{outside} the central region in each galaxy.
As shown in the bottom panels in Figure~\ref{fig:MS}, excluding the central regions reduces the level of galaxy-to-galaxy variations in the mass-weighted median $\Sigmol$ and $\sigmol$.
Nevertheless, the overall trends persist, and the rank correlation \added{of the median $\Sigmol$, $\sigmol$, and $\Pturb$ with all three global galaxy properties} remain significant.

Across the local star-forming galaxy population, we thus conclude that the molecular gas in more massive and actively star-forming galaxies is systematically denser (as traced by $\Sigmol$), more turbulent (as tracked by $\sigmol$ and $\Pturb$), and more strongly self-gravitating (as expressed by $\alphavir$) on ${\sim}100$~pc scales.
We speculate that these trends arise because galaxy global properties correlate with the structural properties on a more \textit{local} scale (e.g., local stellar mass distribution, galaxy dynamical features).
In turn, molecular gas properties on cloud scales are linked to these local structural properties \citep[e.g.,][]{Hughes_etal_2013a,Meidt_etal_2018,Meidt_etal_2020,Schruba_etal_2019,Chevance_etal_2020a,Sun_etal_2020}.
We plan to investigate this topic in more detail in a future round of PHANGS-ALMA analysis.


\section{Summary} \label{sec:summary}

Using the full PHANGS-ALMA \CO21 data set, we measure molecular gas surface density, velocity dispersion, turbulent pressure, and virial parameter on cloud scales in \ngalall\ nearby, massive, star-forming galaxies. We publish the resultant \nlosall\ independent measurements at $150$~pc scales and \nloshresall\ measurements at $90$~pc scales in Table~\ref{tab:product} and summarize their statistics in Table~\ref{tab:stats} and Section~\ref{sec:result:stats}.

Consistent with observations in the PHANGS-ALMA pilot sample \citepalias{Sun_etal_2018} and other galaxies \citep[e.g.,][]{Hughes_etal_2013b,Egusa_etal_2018}, we find that molecular gas properties on \deleted{fixed,} ${\sim}100$~pc scales vary substantially and correlate with location in the host galaxy. Specifically, our key results are:
\begin{enumerate}
    \item Molecular gas surface density, velocity dispersion, and turbulent pressure vary dramatically (by $3.4$, $1.7$, and $6.5$~dex, respectively) across our full sample. The correlation between surface density and velocity dispersion suggests that the gas motions on ${\sim}100$~pc scales are mainly responding to gas self-gravity, though they do also react to external gravity and/or ambient pressure in some regions. The inferred virial parameter has a median value of $2.7$ and a $1\sigma$ range of $0.7$~dex (Figure~\ref{fig:overview} and Section~\ref{sec:result:stats}).
    \item The cloud scale gas surface density, velocity dispersion, and turbulent pressure all increase toward small galactocentric radii, consistent with expectations from vertical dynamical equilibrium and the structure of galaxy disks (Figure~\ref{fig:overview} and Section~\ref{sec:result:rgal}).
    \item The centers of barred galaxies display exceptionally high molecular gas surface densities and velocity dispersions. The high surface densities are likely fueled by gas inflows induced by the stellar bars. The observed excess velocity dispersion at fixed surface density in these regions suggests less bound gas or enhanced bulk flow motions (Figure~\ref{fig:regions} and Section~\ref{sec:result:regions}).
    \item Molecular gas in spiral arm regions shows moderately higher surface densities and appears more turbulent and more bound than the molecular gas detected in the inter-arm regions. This suggests that spiral arms accumulate molecular gas and further mildly alter the gas properties (Figure~\ref{fig:regions} and Section~\ref{sec:result:regions}).
    \item The properties of molecular gas at cloud scale resolution correlate with the properties of the host galaxy. Galaxies with higher stellar mass and more active star formation tend to host molecular gas with higher surface density, higher velocity dispersion, and lower virial parameter (Figure~\ref{fig:regions} and Section~\ref{sec:result:global}).
\end{enumerate}

These observations provide a first comprehensive view of the the properties of molecular gas at cloud scales across the local star-forming galaxy population.
They provide strong evidence that molecular cloud properties are closely coupled to the galactic environment, likely through dynamical processes and stellar feedback.
The empirical relations presented in this work establish the groundwork for unveiling the physics that underpins the molecular cloud--environment connection.

\begin{deluxetable*}{lccccccc}
\tabletypesize{\footnotesize}
\tablecaption{Statistics of Cloud Scale Molecular Gas Properties\label{tab:stats}}
\tablewidth{0pt}
\tablehead{
\colhead{Quantity} &
\colhead{Unit} &
\multicolumn{3}{c}{Area-weighted} &
\multicolumn{3}{c}{Mass-weighted} \\
\cmidrule(lr){3-5} \cmidrule(lr){6-8}
\colhead{} &
\colhead{} &
\colhead{Median} &
\colhead{$1\sigma$ Range (68.3\%)} &
\colhead{$3\sigma$ Range (99.7\%)} &
\colhead{Median} &
\colhead{$1\sigma$ Range (68.3\%)} &
\colhead{$3\sigma$ Range (99.7\%)}
}
\startdata
\multicolumn{8}{c}{Full sample at 150~pc scales (\nlos\ sightlines across \ngal\ galaxies; see Section~\ref{sec:result:stats})} \\
\hline
$I_\mathrm{CO,\,150pc}$ & $\uIco$ & $3.4\times10^0$ & 0.9~dex & 3.2~dex & $1.9\times10^1$ & 1.6~dex & 3.6~dex \\
$\Sigma_\mathrm{mol,\,150pc}$ & $\uSig$ & $2.2\times10^1$ & 0.9~dex & 2.9~dex & $1.1\times10^2$ & 1.5~dex & 3.4~dex \\
$\sigma_\mathrm{mol,\,150pc}$ & $\uV$ & $5.0\times10^0$ & 0.4~dex & 1.7~dex & $9.1\times10^0$ & 0.8~dex & 1.7~dex \\
$P_\mathrm{turb,\,150pc}$ & $\uP$ & $1.8\times10^4$ & 1.6~dex & 6.1~dex & $3.0\times10^5$ & 3.0~dex & 6.5~dex \\
$\alpha_\mathrm{vir,\,150pc}$ & -- & 3.5 & 0.6~dex & 1.9~dex & 2.7 & 0.7~dex & 2.0~dex \\
\hline
\multicolumn{8}{c}{Gas in galaxy disks at 150~pc scales (\nlosdisk\ sightlines across \ngal\ galaxies; see Section~\ref{sec:result:regions} and \ref{sec:result:global})} \\
\hline
$I_\mathrm{CO,\,150pc}$ & $\uIco$ & $3.3\times10^0$ & 0.9~dex & 2.7~dex & $1.2\times10^1$ & 1.1~dex & 3.0~dex \\
$\Sigma_\mathrm{mol,\,150pc}$ & $\uSig$ & $2.1\times10^1$ & 0.9~dex & 2.5~dex & $7.1\times10^1$ & 1.0~dex & 2.8~dex \\
$\sigma_\mathrm{mol,\,150pc}$ & $\uV$ & $4.9\times10^0$ & 0.4~dex & 1.6~dex & $7.5\times10^0$ & 0.5~dex & 1.6~dex \\
$P_\mathrm{turb,\,150pc}$ & $\uP$ & $1.7\times10^4$ & 1.6~dex & 5.4~dex & $1.3\times10^5$ & 1.9~dex & 5.6~dex \\
$\alpha_\mathrm{vir,\,150pc}$ & -- & 3.4 & 0.6~dex & 1.9~dex & 2.7 & 0.7~dex & 2.0~dex \\
\hline
\multicolumn{8}{c}{Gas in the centers of barred galaxies at 150~pc scales (\nlosbar\ sightlines across \ngalbar\ galaxies; see Section~\ref{sec:result:regions})} \\
\hline
$I_\mathrm{CO,\,150pc}$ & $\uIco$ & $6.5\times10^1$ & 1.3~dex & 3.4~dex & $3.0\times10^2$ & 0.9~dex & 2.6~dex \\
$\Sigma_\mathrm{mol,\,150pc}$ & $\uSig$ & $2.8\times10^2$ & 1.3~dex & 3.4~dex & $1.3\times10^3$ & 0.9~dex & 2.6~dex \\
$\sigma_\mathrm{mol,\,150pc}$ & $\uV$ & $2.3\times10^1$ & 0.5~dex & 1.7~dex & $3.4\times10^1$ & 0.4~dex & 1.2~dex \\
$P_\mathrm{turb,\,150pc}$ & $\uP$ & $5.1\times10^6$ & 2.1~dex & 6.5~dex & $5.0\times10^7$ & 1.3~dex & 4.3~dex \\
$\alpha_\mathrm{vir,\,150pc}$ & -- & 6.0 & 0.8~dex & 2.1~dex & 2.7 & 0.7~dex & 2.1~dex \\
\hline
\multicolumn{8}{c}{Full sample at 90~pc scales (\nloshres\ sightlines across \ngalhres\ galaxies; see Section~\ref{sec:result:stats})} \\
\hline
$I_\mathrm{CO,\,90pc}$ & $\uIco$ & $4.2\times10^0$ & 0.9~dex & 2.9~dex & $1.5\times10^1$ & 1.2~dex & 3.3~dex \\
$\Sigma_\mathrm{mol,\,90pc}$ & $\uSig$ & $2.6\times10^1$ & 0.8~dex & 2.6~dex & $8.7\times10^1$ & 1.1~dex & 3.1~dex \\
$\sigma_\mathrm{mol,\,90pc}$ & $\uV$ & $4.5\times10^0$ & 0.4~dex & 1.6~dex & $7.0\times10^0$ & 0.6~dex & 1.7~dex \\
$P_\mathrm{turb,\,90pc}$ & $\uP$ & $2.9\times10^4$ & 1.5~dex & 5.5~dex & $2.3\times10^5$ & 2.2~dex & 6.3~dex \\
$\alpha_\mathrm{vir,\,90pc}$ & -- & 3.8 & 0.6~dex & 1.8~dex & 3.1 & 0.6~dex & 1.8~dex \\
\hline
\multicolumn{8}{c}{Gas in galaxy disks at 90~pc scales (\nlosdiskhres\ sightlines across \ngalhres\ galaxies)} \\
\hline
$I_\mathrm{CO,\,90pc}$ & $\uIco$ & $4.1\times10^0$ & 0.8~dex & 2.6~dex & $1.2\times10^1$ & 1.0~dex & 2.8~dex \\
$\Sigma_\mathrm{mol,\,90pc}$ & $\uSig$ & $2.6\times10^1$ & 0.8~dex & 2.4~dex & $7.1\times10^1$ & 1.0~dex & 2.6~dex \\
$\sigma_\mathrm{mol,\,90pc}$ & $\uV$ & $4.4\times10^0$ & 0.4~dex & 1.5~dex & $6.3\times10^0$ & 0.4~dex & 1.4~dex \\
$P_\mathrm{turb,\,90pc}$ & $\uP$ & $2.7\times10^4$ & 1.5~dex & 5.0~dex & $1.6\times10^5$ & 1.8~dex & 5.0~dex \\
$\alpha_\mathrm{vir,\,90pc}$ & -- & 3.7 & 0.6~dex & 1.8~dex & 2.9 & 0.6~dex & 1.8~dex \\
\enddata
\tablecomments{The area-weighted statistics are derived from percentiles weighted by sightline number counts, whereas the mass-weighted statistics from percentiles weighted by molecular gas mass (equivalent to $\Sigmol$ in our measurement scheme).}
\end{deluxetable*}


\vspace{3em}
{
This work was carried out as part of the PHANGS collaboration.
The work of J.S., A.K.L., and D.U. is partially supported by the National Science Foundation (NSF) under Grants No.\ 1615105, 1615109, and 1653300, as well as by the National Aeronautics and Space Administration (NASA) under ADAP grants NNX16AF48G and NNX17AF39G.
E.S., D.L., T.S., and T.G.W. acknowledge funding from the European Research Council (ERC) under the European Union’s Horizon 2020 research and innovation programme (grant agreement No.\ 694343).
E.R. acknowledges the support of the Natural Sciences and Engineering Research Council of Canada (NSERC), funding reference number RGPIN-2017-03987.
A.U. acknowledges support from the Spanish funding grants AYA2016-79006-P (MINECO/FEDER) and PGC2018-094671-B-I00 (MCIU/AEI/FEDER).
J.M.D.K., M.C, and J.J.K. gratefully acknowledge funding from the DFG through an Emmy Noether Research Group (grant number KR4801/1-1) and the DFG Sachbeihilfe (grant number KR4801/2-1). J.M.D.K. gratefully acknowledges funding from the European Research Council (ERC) under the European Union's Horizon 2020 research and innovation programme via the ERC Starting Grant MUSTANG (grant agreement No.\ 714907).
F.B.\ acknowledges funding from the European Union’s Horizon 2020 research and innovation programme (grant agreement No.\ 726384/EMPIRE).
R.S.K.\ and S.C.O.G.\ acknowledge financial support from the German Research Foundation (DFG) via the collaborative research centre (SFB 881, Project-ID 138713538) `The Milky Way System' (subprojects A1, B1, B2, and B8). They also acknowledge funding from the Heidelberg Cluster of Excellence STRUCTURES in the framework of Germany's Excellence Strategy (grant EXC-2181/1\ -\ 390900948) and from the European Research Council via the ERC Synergy Grant ECOGAL (grant 855130) and the ERC Advanced Grant STARLIGHT (grant 339177).
K.K. gratefully acknowledges funding from the Deutsche Forschungsgemeinschaft (DFG, German Research Foundation) in the form of an Emmy Noether Research Group (grant number KR4598/2-1, PI Kreckel).
A.E.S. is supported by an NSF Astronomy and Astrophysics Postdoctoral Fellowship under award AST-1903834.

This paper makes use of the following ALMA data: \linebreak
ADS/JAO.ALMA\#2012.1.00650.S, \linebreak 
ADS/JAO.ALMA\#2013.1.01161.S, \linebreak 
ADS/JAO.ALMA\#2015.1.00925.S, \linebreak 
ADS/JAO.ALMA\#2015.1.00956.S, \linebreak 
ADS/JAO.ALMA\#2017.1.00392.S, \linebreak 
ADS/JAO.ALMA\#2017.1.00886.L, \linebreak 
ADS/JAO.ALMA\#2018.1.01321.S, \linebreak 
ADS/JAO.ALMA\#2018.1.01651.S. \linebreak 
ALMA is a partnership of ESO (representing its member states), NSF (USA), and NINS (Japan), together with NRC (Canada), NSC and ASIAA (Taiwan), and KASI (Republic of Korea), in cooperation with the Republic of Chile. The Joint ALMA Observatory is operated by ESO, AUI/NRAO, and NAOJ. The National Radio Astronomy Observatory is a facility of the National Science Foundation operated under cooperative agreement by Associated Universities, Inc.

We acknowledge the usage of 
the Extragalactic Distance Database\footnote{\url{http://edd.ifa.hawaii.edu/index.html}} \citep{Tully_etal_2009}
and the SAO/NASA Astrophysics Data System\footnote{\url{http://www.adsabs.harvard.edu}}.

\facilities{ALMA}

\software{
{\tt CASA} \citep{McMullin_etal_2007},
{\tt Astropy} \citep{AstropyCollaboration_etal_2018},
{\tt spectral-cube} \citep{spectral-cube-v0.4.5},
{\tt SpectralCubeTools} (\url{https://github.com/astrojysun/SpectralCubeTools})
}
}


\clearpage
\appendix

\section{Galaxy Sample}\label{apdx:sample}

\renewcommand\thetable{\thesection\arabic{table}}
\setcounter{table}{0}

\startlongtable
\begin{deluxetable*}{lccccrccccccc}
\tabletypesize{\footnotesize}
\tablecaption{Galaxy Sample\label{tab:sample}}
\tablewidth{0pt}
\colnumbers
\tablehead{
\colhead{Galaxy} &
\colhead{Bar} &
\colhead{Arm} &
\colhead{$d$} &
\colhead{$i$} &
\colhead{$\theta_\mathrm{PA}$} &
\colhead{$M_\star$} &
\colhead{SFR} &
\colhead{$\Reff$} &
\colhead{$T_\mathrm{noise,\,150pc}$} &
\colhead{$r_\mathrm{ch,\,150pc}$} &
\colhead{$f_\mathrm{CO,\,150pc}$} &
\colhead{$N_\mathrm{los,\,150pc}$} \\
\colhead{} &
\colhead{} &
\colhead{} &
\colhead{[$\mathrm{Mpc}$]} &
\colhead{[$\deg$]} &
\colhead{[$\deg$]} &
\colhead{[$10^9\,\mathrm{M_\odot}$]} &
\colhead{[$\mathrm{M_\odot/yr}$]} &
\colhead{[$\mathrm{kpc}$]} &
\colhead{[$\mathrm{K}$]} &
\colhead{} &
\colhead{} &
\colhead{}
}
\startdata
Circinus\tablenotemark{$\dagger$} & ? & N & 4.21 & 64.3 & 36.7 & 18.2 & 3.85 & 2.5 & 0.048 & 0.072 & 83\% & 456 \\
IC~1954 & Y & Y & 15.2 & 57.2 & 63.7 & 6.6 & 0.48 & 3.0 & 0.026 & 0.059 & 79\% & 1054 \\
IC~5273 & Y & N & 14.7 & 48.5 & 235.2 & 5.5 & 0.56 & 2.3 & 0.022 & 0.055 & 64\% & 750 \\
NGC~253\tablenotemark{$\dagger$} & Y & N & 3.68 & 75.0 & 52.5 & 38.0 & 4.90 & 4.4 & 0.031 & 0.072 & 88\% & 2203 \\
NGC~300\tablenotemark{$\dagger$} & N & N & 2.08 & 39.8 & 11.4 & 1.7 & 0.14 & 2.2 & 0.011 & 0.123 & 41\% & 127 \\
NGC~628 & N & Y & 9.77 & 8.7 & 20.8 & 18.3 & 1.67 & 4.6 & 0.031 & 0.061 & 83\% & 3239 \\
NGC~685 & Y & N & 16.0 & 32.7 & 99.9 & 7.0 & 0.26 & 4.0 & 0.029 & 0.058 & 41\% & 615 \\
NGC~1087 & Y & N & 14.4 & 40.5 & 357.4 & 6.6 & 1.05 & 3.0 & 0.040 & 0.055 & 75\% & 1165 \\
NGC~1097 & Y & Y & 14.2 & 48.6 & 122.8 & 60.8 & 5.08 & 5.4 & 0.032 & 0.062 & 85\% & 3093 \\
NGC~1300 & Y & Y & 26.1 & 31.8 & 276.9 & 71.9 & 2.06 & 9.1 & 0.096 & 0.054 & 48\% & 1037 \\
NGC~1317 & Y & N & 19.0 & 24.5 & 221.5 & 36.6 & 0.40 & 4.4 & 0.032 & 0.063 & 105\%\tablenotemark{$\star$} & 575 \\
NGC~1365 & Y & Y & 18.1 & 55.4 & 202.4 & 66.8 & 14.34 & 11.8 & 0.067 & 0.191 & 88\% & 2073 \\
NGC~1385 & ? & Y & 22.7 & 45.4 & 179.6 & 16.6 & 3.50 & 4.9 & 0.072 & 0.054 & 67\% & 1796 \\
NGC~1433 & Y & N & 16.8 & 28.6 & 198.0 & 52.9 & 0.81 & 8.3 & 0.057 & 0.055 & 58\% & 684 \\
NGC~1511 & ? & N & 15.6 & 73.5 & 296.9 & 7.6 & 2.27 & 2.8 & 0.038 & 0.063 & 89\% & 778 \\
NGC~1512 & Y & Y & 16.8 & 42.5 & 263.8 & 38.3 & 0.91 & 7.2 & 0.052 & 0.057 & 61\% & 689 \\
NGC~1546 & N & N & 18.0 & 70.1 & 147.8 & 22.8 & 0.80 & 3.2 & 0.030 & 0.057 & 97\% & 972 \\
NGC~1559 & Y & N & 19.8 & 58.7 & 245.9 & 21.3 & 3.72 & 3.5 & 0.056 & 0.056 & 75\% & 2218 \\
NGC~1566 & Y & Y & 18.0 & 30.5 & 216.5 & 53.3 & 4.49 & 8.4 & 0.057 & 0.058 & 97\% & 3944 \\
NGC~1637 & Y & Y & 9.77 & 31.1 & 20.6 & 7.7 & 0.66 & 1.1 & 0.012 & 0.054 & 91\% & 1360 \\
NGC~1672 & Y & Y & 11.9 & 43.8 & 135.9 & 17.7 & 2.73 & 5.1 & 0.052 & 0.064 & 82\% & 1291 \\
NGC~1792 & N & N & 12.8 & 64.7 & 318.9 & 23.3 & 2.21 & 3.2 & 0.028 & 0.066 & 94\% & 1468 \\
NGC~2090 & N & Y & 11.8 & 64.4 & 192.4 & 11.1 & 0.32 & 2.5 & 0.042 & 0.061 & 80\% & 516 \\
NGC~2283 & Y & Y & 10.4 & 44.2 & 356.2 & 3.6 & 0.26 & 2.1 & 0.036 & 0.061 & 44\% & 287 \\
NGC~2566 & Y & Y & 23.7 & 48.5 & 312.0 & 40.6 & 8.47 & 5.7 & 0.072 & 0.064 & 79\% & 1978 \\
NGC~2835 & Y & Y & 10.1 & 41.1 & 0.2 & 5.9 & 0.76 & 2.8 & 0.056 & 0.060 & 28\% & 182 \\
NGC~2903 & Y & N & 8.47 & 67.0 & 205.4 & 28.9 & 2.08 & 4.5 & 0.026 & 0.065 & 90\% & 2390 \\
NGC~2997 & ? & Y & 11.3 & 31.9 & 109.3 & 31.2 & 2.79 & 5.0 & 0.026 & 0.063 & 86\% & 5380 \\
NGC~3137 & ? & N & 14.9 & 70.1 & 358.9 & 5.8 & 0.41 & 4.6 & 0.033 & 0.056 & 70\% & 488 \\
NGC~3351 & Y & N & 10.0 & 45.1 & 193.2 & 20.8 & 1.09 & 3.1 & 0.039 & 0.062 & 74\% & 991 \\
NGC~3507 & Y & Y & 20.9 & 24.2 & 55.6 & 27.3 & 0.75 & 3.5 & 0.067 & 0.060 & 45\% & 1090 \\
NGC~3511 & Y & N & 9.95 & 75.0 & 256.7 & 5.1 & 0.42 & 3.0 & 0.020 & 0.058 & 87\% & 769 \\
NGC~3521 & N & N & 11.2 & 69.0 & 343.0 & 66.3 & 2.59 & 5.6 & 0.023 & 0.056 & 90\% & 3770 \\
NGC~3596 & N & N & 10.1 & 21.6 & 78.1 & 3.5 & 0.23 & 1.7 & 0.052 & 0.060 & 72\% & 495 \\
NGC~3621 & N & N & 6.56 & 65.4 & 343.8 & 9.2 & 0.79 & 2.9 & 0.013 & 0.063 & 91\% & 1487 \\
NGC~3626 & Y & N & 20.0 & 46.6 & 165.2 & 27.5 & 0.23 & 3.3 & 0.084 & 0.057 & 57\% & 150 \\
NGC~3627 & Y & Y & 10.57 & 56.5 & 174.0 & 53.1 & 3.24 & 5.2 & 0.033 & 0.061 & 89\% & 2933 \\
NGC~4207\tablenotemark{$\ddagger$} & ? & N & 16.8 & 62.5 & 120.5 & 5.1 & 0.22 & 1.3 & 0.062 & 0.067 & 91\% & 147 \\
NGC~4254 & N & Y & 16.8 & 35.3 & 68.5 & 37.8 & 4.95 & 3.6 & 0.053 & 0.056 & 84\% & 6438 \\
NGC~4293 & Y & N & 16.0 & 65.0 & 48.3 & 30.6 & 0.60 & 3.8 & 0.061 & 0.075 & 81\% & 164 \\
NGC~4298 & N & N & 16.8 & 59.6 & 314.1 & 13.0 & 0.56 & 2.7 & 0.025 & 0.056 & 93\% & 2328 \\
NGC~4303 & Y & Y & 17.6 & 20.0 & 310.6 & 50.4 & 5.63 & 6.2 & 0.066 & 0.061 & 82\% & 3945 \\
NGC~4321 & Y & Y & 15.2 & 39.1 & 157.7 & 49.4 & 3.41 & 6.2 & 0.058 & 0.058 & 77\% & 4923 \\
NGC~4424\tablenotemark{$\ddagger$} & ? & N & 16.4 & 58.2 & 88.3 & 8.3 & 0.31 & 3.3 & 0.060 & 0.071 & 103\%\tablenotemark{$\star$} & 123 \\
NGC~4457 & Y & N & 15.6 & 17.4 & 78.7 & 25.7 & 0.34 & 3.1 & 0.041 & 0.060 & 93\% & 645 \\
NGC~4496A & Y & N & 14.9 & 55.3 & 49.7 & 4.2 & 0.61 & 3.1 & 0.057 & 0.058 & 29\% & 168 \\
NGC~4535 & Y & Y & 15.8 & 42.1 & 179.3 & 32.3 & 2.07 & 5.8 & 0.053 & 0.059 & 75\% & 2433 \\
NGC~4536 & Y & Y & 15.2 & 64.8 & 307.4 & 20.0 & 2.99 & 4.2 & 0.025 & 0.059 & 88\% & 2025 \\
NGC~4540 & Y & N & 16.8 & 38.3 & 14.3 & 6.8 & 0.19 & 1.8 & 0.059 & 0.063 & 65\% & 428 \\
NGC~4548 & Y & Y & 16.2 & 38.3 & 138.0 & 45.6 & 0.53 & 5.1 & 0.035 & 0.060 & 49\% & 1027 \\
NGC~4569 & Y & N & 16.8 & 70.0 & 18.0 & 67.2 & 1.54 & 8.9 & 0.038 & 0.058 & 85\% & 2544 \\
NGC~4571 & N & N & 14.9 & 31.9 & 217.4 & 11.6 & 0.30 & 3.3 & 0.058 & 0.059 & 42\% & 711 \\
NGC~4579 & Y & Y & 16.8 & 37.3 & 92.5 & 83.1 & 1.08 & 5.7 & 0.039 & 0.057 & 70\% & 3078 \\
NGC~4689 & N & N & 16.8 & 39.0 & 164.3 & 17.0 & 0.52 & 4.2 & 0.060 & 0.058 & 72\% & 1827 \\
NGC~4694\tablenotemark{$\ddagger$} & N & N & 16.8 & 60.7 & 143.3 & 7.8 & 0.15 & 3.0 & 0.055 & 0.056 & 38\% & 76 \\
NGC~4731 & Y & Y & 12.4 & 64.0 & 255.4 & 3.3 & 0.42 & 4.0 & 0.017 & 0.052 & 56\% & 261 \\
NGC~4781 & Y & N & 15.3 & 56.4 & 288.1 & 8.0 & 0.84 & 2.4 & 0.022 & 0.055 & 79\% & 1411 \\
NGC~4826\tablenotemark{$\ddagger$} & N & N & 4.36 & 58.6 & 293.9 & 16.0 & 0.20 & 1.7 & 0.014 & 0.068 & 97\% & 147 \\
NGC~4941 & ? & N & 14.0 & 53.1 & 202.6 & 12.4 & 0.36 & 3.4 & 0.020 & 0.054 & 80\% & 1196 \\
NGC~4951 & N & N & 12.0 & 70.5 & 92.0 & 3.9 & 0.21 & 2.5 & 0.034 & 0.063 & 71\% & 214 \\
NGC~5042 & ? & N & 12.6 & 51.4 & 190.1 & 4.7 & 0.33 & 2.9 & 0.035 & 0.056 & 35\% & 300 \\
NGC~5068 & Y & N & 5.16 & 27.0 & 349.0 & 2.2 & 0.28 & 2.1 & 0.037 & 0.065 & 46\% & 222 \\
NGC~5134 & Y & N & 18.5 & 22.7 & 311.6 & 21.6 & 0.37 & 4.2 & 0.047 & 0.059 & 59\% & 538 \\
NGC~5248 & ? & Y & 12.7 & 49.5 & 106.2 & 17.0 & 1.54 & 3.2 & 0.049 & 0.065 & 87\% & 1190 \\
NGC~5530 & ? & N & 11.8 & 61.9 & 305.4 & 9.4 & 0.31 & 2.8 & 0.046 & 0.057 & 70\% & 798 \\
NGC~5643 & Y & Y & 11.8 & 29.9 & 318.7 & 18.2 & 2.14 & 3.5 & 0.034 & 0.058 & 83\% & 2667 \\
NGC~6300 & Y & N & 13.1 & 49.3 & 105.5 & 29.2 & 2.39 & 4.0 & 0.050 & 0.059 & 81\% & 2120 \\
NGC~6744 & Y & Y & 11.6 & 53.2 & 14.3 & 48.8 & 2.28 & 9.7 & 0.065 & 0.065 & 66\% & 2511 \\
NGC~7456 & ? & N & 7.94 & 63.7 & 12.9 & 1.2 & 0.06 & 2.2 & 0.011 & 0.057 & 48\% & 133 \\
NGC~7496 & Y & N & 18.7 & 34.7 & 196.4 & 9.8 & 2.16 & 3.3 & 0.029 & 0.056 & 79\% & 1557 \\
\enddata
\tablecomments{
(2--3) If the galaxy has identifiable stellar bars and spiral arms (? -- ambiguous; see M.~Querejeta et al. 2020, in preparation);
(4) Distance \citep{Tully_etal_2009};
(5--6) Galaxy inclination and position angle \citep{Lang_etal_2020};
(7--9) Galaxy global stellar mass, SFR, and the effective (half-mass) radius estimated from the measured stellar scale length \citep[A.~K.~Leroy et al. 2020a, in preparation]{Leroy_etal_2019};
(10--11) CO data rms noise and channel-to-channel correlation at $150$~pc resolution;
(12) CO flux completeness at $150$~pc resolution;
(13) Number of independent sightlines at $150$~pc resolution.\\
(A machine readable version of this table is available in the \href{https://www.canfar.net/storage/list/phangs/RELEASES/Sun_etal_2020b}{PHANGS CADC storage}.)
}
\tablenotetext{\dagger}{These three very nearby galaxies are only observed by the ACA 7-m and total power telescopes. Because of their proximity, the data still have linear resolutions matched to the other galaxies in the sample.}
\tablenotetext{\ddagger}{Measurements in these four galaxies are not presented in Section~\ref{sec:result}.}
\tablenotetext{\star}{The estimated CO flux completeness exceeds $100$\% for these two targets. This is due to either low S/N data (NGC~4424) or calibration mismatch between the interferometric and single dish data (NGC~1317).}
\end{deluxetable*}

\clearpage

\section{Data Product}\label{apdx:product}

\renewcommand\thetable{\thesection\arabic{table}}
\setcounter{table}{0}

\startlongtable
\begin{deluxetable*}{ccccccccccc}
\tabletypesize{\footnotesize}
\tablecaption{Table of Key Measurements\label{tab:product}}
\tablewidth{0pt}
\colnumbers
\tablehead{
\colhead{Galaxy} &
\colhead{Resolution} &
\colhead{$\rgal$} &
\colhead{Center} &
\colhead{Arm} &
\colhead{Inter-arm} &
\colhead{$\I\sbsc{CO(2\text{-}1)}$} &
\colhead{$\Sigmol$} &
\colhead{$\sigmol$} &
\colhead{$\Pturb$} &
\colhead{$\alphavir$} \\
\colhead{} &
\colhead{[$\mathrm{pc}$]} &
\colhead{[$\mathrm{kpc}$]} &
\colhead{} &
\colhead{} &
\colhead{} &
\colhead{[$\uIco$]} &
\colhead{[$\uSig$]} &
\colhead{[$\uV$]} &
\colhead{[$\uP$]} &
\colhead{}
}
\startdata
Circinus & 150 & 0.000 & 1 & 0 & 0 & 7.680e+02 & 3.423e+03 & 7.664e+01 & 6.574e+08 & 5.280e+00 \\
Circinus & 150 & 0.154 & 1 & 0 & 0 & 4.755e+02 & 2.161e+03 & 4.053e+01 & 1.160e+08 & 2.339e+00 \\
Circinus & 150 & 0.154 & 1 & 0 & 0 & 3.649e+02 & 1.659e+03 & 4.124e+01 & 9.228e+07 & 3.154e+00 \\
Circinus & 150 & 0.290 & 1 & 0 & 0 & 3.433e+02 & 1.596e+03 & 4.595e+01 & 1.101e+08 & 4.071e+00 \\
Circinus & 150 & 0.290 & 1 & 0 & 0 & 5.191e+02 & 2.398e+03 & 7.411e+01 & 4.305e+08 & 7.048e+00 \\
Circinus & 150 & 0.307 & 1 & 0 & 0 & 2.265e+02 & 1.053e+03 & 2.411e+01 & 2.001e+07 & 1.698e+00 \\
Circinus & 150 & 0.307 & 1 & 0 & 0 & 2.698e+02 & 1.252e+03 & 2.277e+01 & 2.121e+07 & 1.275e+00 \\
Circinus & 150 & 0.322 & 1 & 0 & 0 & 3.209e+02 & 1.493e+03 & 4.636e+01 & 1.049e+08 & 4.430e+00 \\
Circinus & 150 & 0.322 & 1 & 0 & 0 & 3.470e+02 & 1.624e+03 & 5.172e+01 & 1.420e+08 & 5.068e+00 \\
Circinus & 150 & 0.334 & 1 & 0 & 0 & 1.989e+02 & 9.334e+02 & 3.261e+01 & 3.244e+07 & 3.505e+00 \\
... & ... & ... & ... & ... & ... & ... & ... & ... & ... & ... \\
\enddata
\tablecomments{
(This table is available in its entirety  in the \href{https://www.canfar.net/storage/list/phangs/RELEASES/Sun_etal_2020b}{PHANGS CADC storage}.)
}
\end{deluxetable*}

\clearpage

\section{Data Censoring Function}\label{apdx:censoring}

\setcounter{equation}{0}

As mentioned in Section~\ref{sec:method}, our data cube masking scheme introduces a censoring effect that excludes sightlines with low $\ICO$ and high $\sigCO$.
Here we provide the analytic expression for this censoring function.

We consider a generic masking scheme requiring $N$ consecutive channels with $\text{S/N}>X\sbsc{th}$.
The intrinsic CO line profile is assumed to be Gaussian, with a peak brightness temperature of $T\sbsc{peak}$ and a 1$\sigma$ line width of $\sigCO$.
We also assume the line peak is located right at the center of the $N$ consecutive channels, each of which has a channel width $v\sbsc{ch}$.
If the CO intensity in the ``edge'' channels (i.e., $\pm N/2$ channels away from the line center) exceeds $X\sbsc{th}\,T\sbsc{noise}\,v\sbsc{ch}$, then all channels in between also exceed this threshold, and thus this CO line should enter the signal mask.
Following this argument, we can get an expression for the censoring function by integrating the line profile within that ``edge'' channel:
\begin{equation}
    \int^{(N/2)\,v\sbsc{ch}}_{(N/2-1)\,v\sbsc{ch}} T\sbsc{peak}\exp(-v^2/2\sigCO^2)\,\mathrm{d}v > X\sbsc{th}\,T\sbsc{noise}\,v\sbsc{ch}~ \label{eq:threshold}.
\end{equation}
\noindent Recasting this integral by the error function and re-expressing $T\sbsc{peak}$ with line-integrated intensity $\ICO=\sqrt{2\pi}\,T\sbsc{peak}\,\sigCO$, we have
\begin{align}
    \ICO >\;&\frac{1}{2}X\sbsc{th}\,T\sbsc{noise}\,v\sbsc{ch} \nonumber\\ &\left[\text{erf}\!\left(\frac{N}{2}\frac{v\sbsc{ch}}{\sqrt{2}\,\sigCO}\right) - \text{erf}\!\left(\frac{N-2}{2}\frac{v\sbsc{ch}}{\sqrt{2}\,\sigCO}\right)\right]^{-1}~. \label{eq:censoring}
\end{align}

The above derivation assumes an infinitely sharp spectral response curve, which is inconsistent with the non-zero channel-to-channel correlation estimated for our data (see Table~\ref{tab:sample}).
To address this, we introduce a three-element Hann kernel of the shape $[k,\;1-2k,\;k]$ to model the spectral response curve. Here the value $k$ is determined so that the resultant channel-to-channel correlation matches the estimated $r\sbsc{ch}$ for our data \citep[following equation~15 in][]{Leroy_etal_2016}.
Convolving the left hand side of Equation~\ref{eq:threshold} with this kernel and recasting the formula into a similar form as Equation~\ref{eq:censoring}, we get a modified censoring function that accounts for the realistic spectral response curve\footnote{The \texttt{Python} realization of this censoring function calculation is available at: \url{https://github.com/astrojysun/SpectralCubeTools}.}:
\begin{align}
    \ICO >\;&\frac{1}{2}X\sbsc{th}\,T\sbsc{noise}\,v\sbsc{ch} \nonumber\\
    &\left[
    k\cdot\text{erf}\!\left(\frac{N+2}{2}\frac{v\sbsc{ch}}{\sqrt{2}\,\sigCO}\right) + \right. \nonumber\\
    &\;\;\;(1-3k)\cdot\text{erf}\!\left(\frac{N}{2}\frac{v\sbsc{ch}}{\sqrt{2}\,\sigCO}\right) - \nonumber\\
    &\;\;\;(1-3k)\cdot\text{erf}\!\left(\frac{N-2}{2}\frac{v\sbsc{ch}}{\sqrt{2}\,\sigCO}\right) - \nonumber\\
    &\;\;\;\left.k\cdot\text{erf}\!\left(\frac{N-4}{2}\frac{v\sbsc{ch}}{\sqrt{2}\,\sigCO}\right)
    \right]^{-1}~. \label{eq:real-censoring}
\end{align}
Taking $N=2$, $X\sbsc{th}=2$, $v\sbsc{ch}=2.5\,\uV$, and the corresponding $T\sbsc{noise}$ and $k$ values for each galaxy in Equation~\ref{eq:real-censoring}, we recover the censoring function shown in Figure~\ref{fig:overview}.


\bibliography{main.bib}


\end{document}